\documentstyle[graphicx,amsmath,amssymb]{jaa}

\begin{document}
\title
{Radio relics in cosmological simulations}
\author[Hoeft et al.] 
{
M. Hoeft\thanks{e-mail: hoeft@tls-tautenburg.de} \\
 Th\"uringer Landessternwarte, Sternwarte 5, 07778, Tautenburg, Germany
  \and
S. E. Nuza \&  S. Gottl\"ober \\
  Leibniz-Institut f\"ur Astrophysik Potsdam, An der Sternwarte 16, \\ 14482 Potsdam, Germany 
  \and
R. J. van Weeren \&  H. J. A. R\"ottgering \\
  Sterrewacht Leiden, PO Box 9513, 2300 RA Leiden, The Netherlands
  \and
M. Br\"uggen \\
  Jacobs University Bremen, Campus Ring 1, 28725 Bremen, Germany
}

\pubyear{}
\volume{}
\date{Received xxx; accepted xxx}
\maketitle
\label{firstpage}

\begin{abstract}

  Radio relics have been discovered in many galaxy clusters. They are believed
  to trace shock fronts induced by cluster mergers. Cosmological simulations
  allow to study merger shocks in detail since the intra-cluster medium is
  heated by shock dissipation. Using high resolution cosmological simulations,
  identifying shock fronts and applying a parametric model for the radio
  emission allows us to simulate the formation of radio relics. We analyze a
  simulated shock front in detail. We find a rather broad Mach number
  distribution. The Mach number affects strongly the number density of
  relativistic electrons in the downstream area, hence, the radio luminosity
  varies significantly across the shock surface. The abundance of radio relics
  can be modeled with the help of the {\it radio power probability distribution}
  which aims at predicting radio relic number counts. Since the actual electron
  acceleration efficiency is not known, predictions for the number counts need
  to be normalized by the observed number of radio relics. For the
  characteristics of upcoming low frequency surveys we find that about thousand
  relics are awaiting discovery.

\end{abstract}


\section{Introduction}

  % Introduce radio relics
  In many galaxy clusters diffuse radio emission has been discovered in the
  periphery of the cluster. If the emission has no optical counterpart, i.e. it
  is apparently not related to an active galaxy, the diffuse emission is
  classified as a `radio relic'. Initially it was assumed that they corresponded
  to the left-overs from former AGN activity, justifying the name `relic'.
  Ensslin et al. (1998) instead suggested that they trace merger shock fronts in
  the intra-cluster medium (ICM). Within the past years several spectacular
  relics have been discovered, e.g. the very bright relic in Abell~3667
  (R\"ottgering et al. 1997), the almost ring-like structure in Abell~3376
  (Bagchi et al. 2006), the about 2~Mpc long but exceptionally narrow relic in
  CIZA~2242 (van Weeren et al. 2010), and recently a double relic in
  a cluster which has been discovered by its Sunyaev-Zeldovich signature, PLCK
  G287.0+32.9 (Bagchi et al. 2011).

  % Shock fronts
  All mentioned radio relics have morphologies similar to shock fronts in
  cluster simulations (e.g. Paul et al. 2011). Typically, relics are aligned
  with X-ray isophotes. For a few clusters a shock front in the ICM has been
  identified in X-ray observations (Markevitch 2010). For instance, for A~3667
  the front coincides with the outer edge of the north-west relic. Moreover, the
  temperature jump has been measured. Rankine-Hugoniot jump conditions for
  hydrodynamical shock fronts allow us to estimate the Mach number of the shock.
  Since the ICM temperature is correlated with the mass of the cluster, which
  also determines the merger velocity, Mach numbers of merger shocks are in the
  range of 2 to 4, at maximum.

  % Introduce cosmological simulations
  Cosmological simulations are ideal to study the formation of merger shocks.
  Since the clusters are heated by the shock fronts any hydrodynamical
  simulation needs to include shock dissipation. Ryu et al. (2003) introduced
  the distinction between `accretion shocks', where photo-ionized gas is
  shock-heated for the first time and `merger shocks', where the upstream gas
  has already encountered earlier phases of shock-heating. This clearly limits
  the Mach numbers of merger shocks. Even if cosmological hydrodynamical codes
  cover properly the shock dissipation, they do usually not determine the Mach
  number explicitly. For Smoothed-Particle Hydrodynamics (SPH) Pfrommer et al.
  (2006) developed a method based on the entropy gain of the particles. Also for
  SPH, we developed a method based on the entropy gradients (Hoeft et al. 2008).
  Skillman et al. (2010) used a shock detection method for Eulerian simulations
  to study radio relics. Recently, Vazza et al. (2009) developed a method to
  increase the level of refinement in adaptive-mesh refinement simulations at
  the locations of shock fronts. Hence, even if merger shocks are basically
  covered in any hydrodynamical cosmological simulation, locating the shocks,
  determining the Mach number, and achieving sufficient numerical resolution is
  still a challenge.

  % Constraints on simulating relics
  A clear relic example has been found in CIZA~2242 (van Weeren et al. 2010).
  First, it has an overall spectrum which is close to a power-law and has a
  slope of $-1.08 \pm 0.05$. Secondly, the clear spectral gradient across the
  relic supports that electrons get accelerated at the front and then lose
  energy while moving away from the front. Finally, the spectral aging of the
  electrons is imprinted as spectral steeping in the radio emission. This
  summarizes what is currently widely believed as the formation scenario for
  relics: At the shock fronts electrons get accelerated via diffusive shock
  acceleration (DSA, Drury 1983; Blandford \& Eichler, 1987; Malkov \& Drury,
  2001), with a maximal slope in the radio spectrum of -0.5. The relativistic
  electrons advect downstream and cool, hence the spectrum steepens with
  increasing distance from the shock front. However, alternative scenarios has
  been suggested, see e.g. Keshet (2011).
  \\

  % In this paper
  In this paper we present current efforts to model radio relics in
  cosmological simulations. First, we summarize how to model radio relics.
  Secondly, we present results from a recent high resolution simulation.
  Finally, we briefly introduce predictions for upcoming low-frequency
  radio surveys.

\section{Modeling the radio emission of relics}

  % Shock fronts as average
  Apparently, radio relics in galaxy clusters are closely related to merger shock
  fronts. In a plausible scenario for the formation of the radio relics a small
  fraction of thermal electrons gets accelerated at the front and the resulting
  relativistic electrons emit synchrotron radiation while advecting downstream.
  Based on this scenario a radio relic model to be implemented in cosmological
  simulations can be decomposed into two parts, namely electron acceleration at
  the shock front and downstream synchrotron emission.

  % Shock front and electron acceleration
  Given the plasma conditions of the ICM in the periphery of galaxy clusters,
  namely electron densities $ \sim 10^{-4} \: \rm cm^{-3}$, temperatures of a
  few keV, and magnetic fields of the order of $ 1 ~ \rm \mu G$, the shock
  fronts are collisionless, i.e. the energy dissipation cannot be mediated via
  particle collisions. For rather low Mach numbers anomalous resistivity may be
  sufficient to dissipate the kinetic energy of the upstream flow. When the Mach
  number is above the critical Mach number of 2.76 proton reflection at the
  shock front may cause the necessary dissipation. The detailed processes at the
  shock front are complex, see e.g. Treumann (2009) for a review of collisonless
  shocks. The characteristic length scale for the shock transition region is the
  proton gyroradius. For thermal protons in the periphery of galaxy clusters
  this radius is of the order of 1\:cm, i.e. even less than the mean particle
  separation. Electrons may get accelerated first via electric fields inherent
  to the shock structure and higher energies may be achieved via diffusive shock
  acceleration. In any case, radio relics -observationally and in simulations-
  represent an averaging over a huge shock area, compared to the characteristic
  shock transition scale, namely the proton gyroradius.

\begin{figure}
\centering
\includegraphics[width=0.96\textwidth]{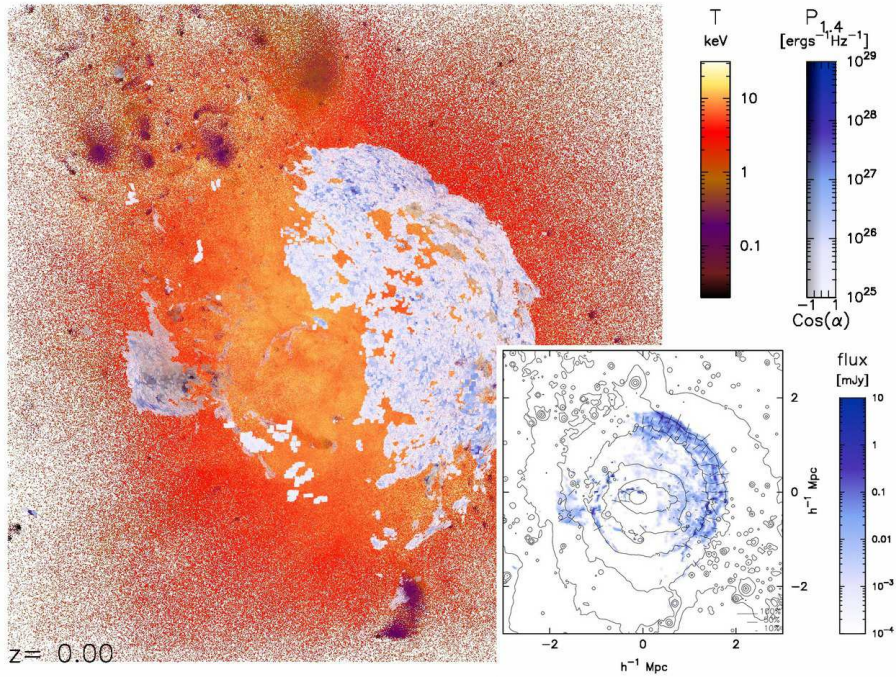}
\caption{
   Visualization of the shock front: Particles in the simulation which have been
   identified as being part of a shock depicted with small white/blue squares.
   Other particles and those for which the Mach number is very low are depicted
   as brown/red/yellow dots. The shock fronts has roughly a shape of a shell
   segment. However, there are many small scale variations in shape and radio
   luminosity. The lower right inset shows the projected X-ray (contours) and
   radio emission.   
   }
\label{fig:shock-visual}
\end{figure}

  % Model in HB07
  Including the detailed physics of the electron acceleration at the shock front
  is beyond the scope of any current simulation of galaxy cluster mergers.
  Instead, the resulting distribution of relativistic electrons has to be
  implemented as a parametric subgrid model. In Hoeft \& Br\"uggen (2007) we
  constructed the following model: The {\it slope} of the average electron
  population is given by the predictions of DSA in the test-particle regime
  (Drury 1983; Blandford \& Eichler, 1987; Malkov \& Drury, 2001). We assume
  that this reflects the electron spectrum averaged over a resolution element in
  the simulation. Moreover, we assumed that a fraction $\xi_{\rm e}$ of the
  energy dissipated at the shock front is transferred to the supra-thermal
  electrons. This allows us to normalize the electron spectrum. However, a lower
  energy cut-off, $E_{\rm low}$, is needed for that. We argued that the energy
  fraction $\xi_{\rm e}$ refers to any energy above the thermal pool, hence
  $E_{\rm low}$ is the energy where the power law distribution meets the thermal
  one. Based on this condition $E_{\rm low}$ can be determined for given gas
  temperature, $\xi_{\rm e}$ and slope of the electron distribution. In our
  model we have also defined a function $\Psi({\cal M})$ which reflects the
  dependence of the radio emission as a function of the Mach number. We find
  that the Mach number needs to be above 2-3 to cause a significant amount of
  radio emission.

  % Assumptions for B
  The initial electron spectrum `ages' while the plasma moves away from the
  shock front. Electrons in the energy range relevant for the radio emission
  cool due to synchrotron emission and inverse Compton (IC) losses. Hence the
  evolution of the spectrum can be modeled at least numerically. However, we can
  only speculate about the strength and the structure of magnetic fields in the
  periphery of galaxy clusters. The upper limit for IC emission in the northwest
  relic in A~3667 puts a lower limit on the magnetic field in that region,
  namely $\gtrsim 2 ~ \mu \rm G$ (Nakazawa et al. 2009). The narrowness of the
  relic in CIZA~2242 indicates that the magnetic field should be $\gtrsim 5 ~ \mu
  \rm G$, if we exclude the $B < 1 ~ \mu \rm G$ solution (van Weeren et al. 2010).
  Rotation Measure analysis of the magnetic field in the Coma cluster indicates
  that the magnetic field scales with the electron density as $ B = 4.7 ~ \mu {\rm
  G} \: ( \: n_{\rm e} \: /  \: 3.44 \times 10^{-3} \:\rm cm^{-3} \: )^{0.5} $ (Bonafede et
  al. 2010).

  % Use a subgrid model
  The width of the relic in CIZA\:2242 shows that the cooling of the electron
  population takes place at maximum on scales of the order of 10\:kpc. Hence to
  resolve the cooling in the simulation in the relic region a numerical
  resolution significantly below 10\:kpc is needed. For a typical shock speed of
  $ 2000 ~ \rm km \, s^{-1}$ this implies a time resolution of $\sim 5 ~ \rm
  Myr$. Moreover, cooling of the relativistic electrons takes place on even
  shorter time scales. Since the resolution in our simulations is $\gtrsim 10 ~ \rm
  kpc$ we treat the radio emission entirely as a subgrid model. We compute the
  radio power per shock surface given the Mach number and the downstream
  properties.

\section{Emission maps in a cosmological simulation}

\begin{figure}
\centering
\includegraphics[width=0.49 \textwidth]{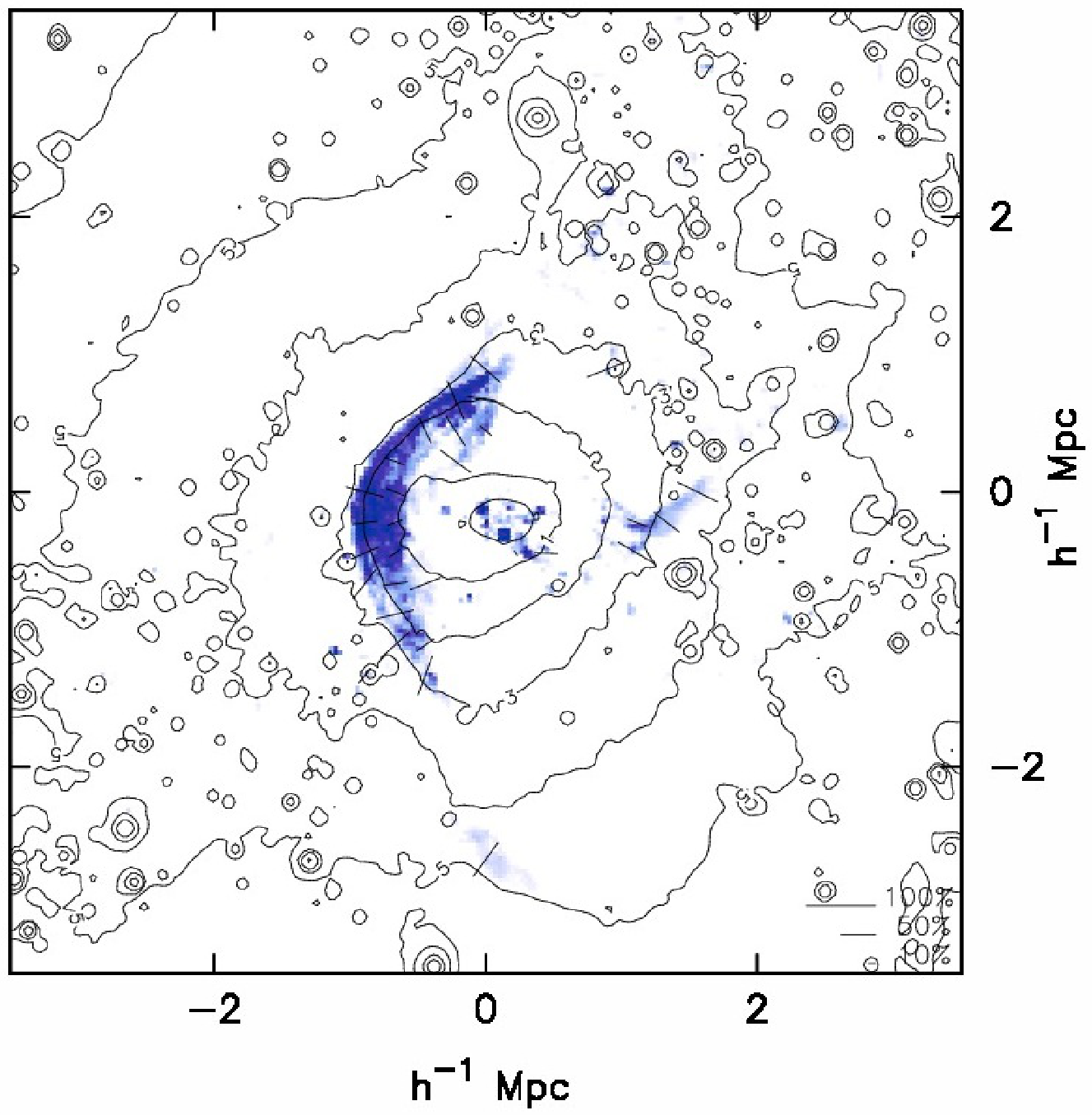}
\hfill
\includegraphics[width=0.49 \textwidth]{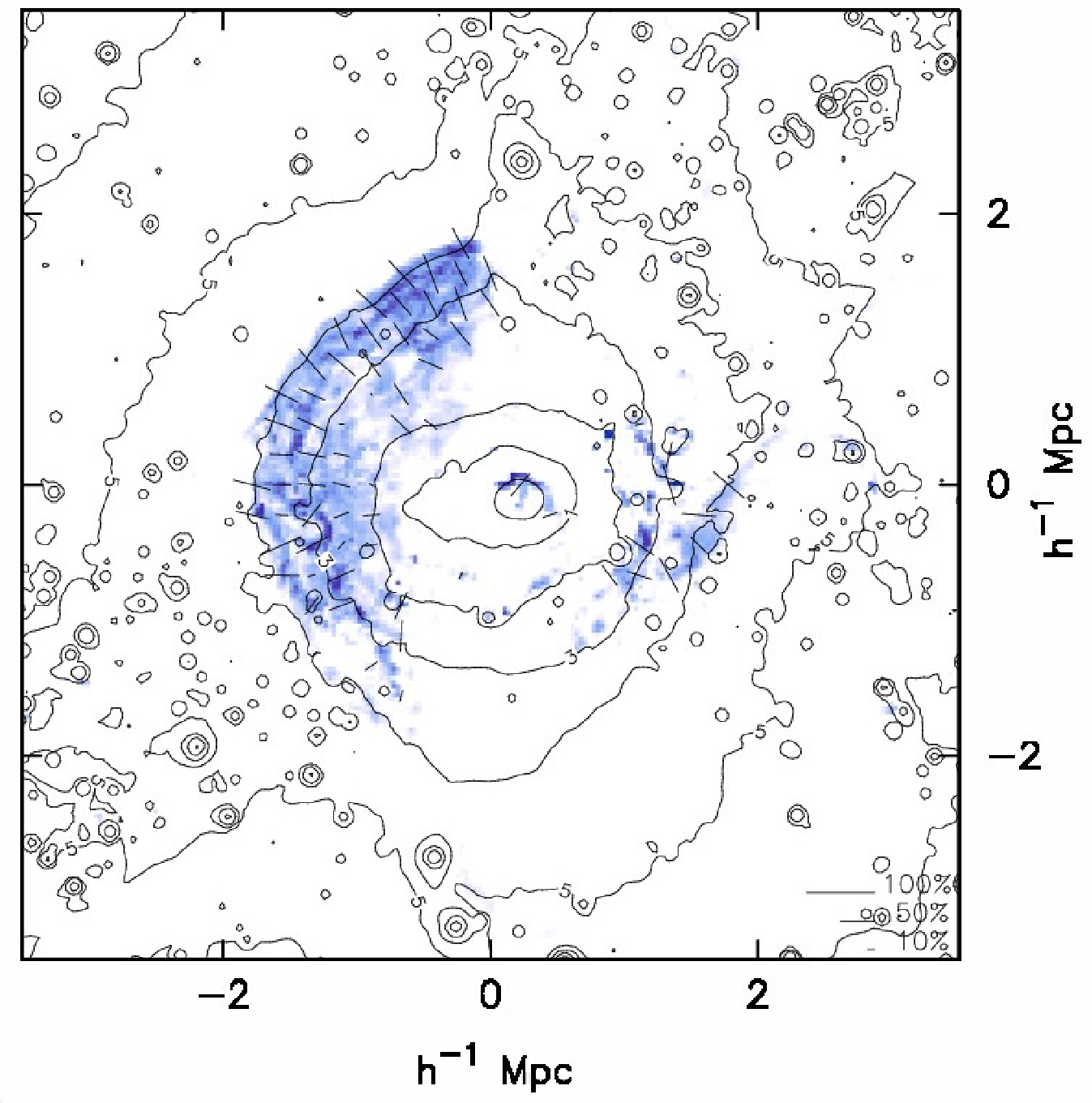}
\caption{
  Evolution of the radio relic caused by a merger shock front. One can clearly
  see how the surface brightness of the relics fades from the early state (left
  panel) to the later state (right panel, about 300~Myr later), even if the Mach
  number actually increases.
  }
\label{fig:evolution}
\end{figure}

  % resimulated clusters
  We aim to study the formation of radio relics in realistic cosmological
  scenarios. We simulate the cluster formation using the TreeSPH code {\sc
  Gadget} (Springel 2005), including hydrodynamics but no radiative cooling or
  heating processes. To obtain sufficiently massive clusters we use a simulation
  box with a side length of $160 ~ h^{-1} \: \rm Mpc$ and with an initial
  resolution of $256^3$ particles. To increase the effective resolution we apply
  the zoom-in technique achieving an effective resolution of $2048^3$. At the
  location of radio relics the overdensity is typically of the order of $10^3$.
  With 64 particles in the SPH kernel the resulting hydrodynamical resolution is
  about  $ 30 ~ h^{-1} \: \rm kpc $. Therefore, as stated above, treating the
  downstream aging of electrons accelerated at the shock front entirely as a
  subgrid model is appropriate.

  % Finding shock fronts
  The ICM medium is heated by the energy dissipation of accretion and merger
  shock fronts. Hence, any simulation of the ICM needs to account for shock
  dissipation, but in SPH the Mach number is not explicitly determined. For our
  radio emission subgrid model we need both, the Mach number and the surface
  area of the shock front. In Hoeft et al. (2008) we have presented a scheme for
  determining the Mach number in SPH simulations. In a slightly revised version
  of the scheme we determine candidate shock fronts using the pressure gradient.
  A conservative estimate for the Mach number is obtained from the velocity
  divergence and density and entropy jumps (see Nuza et al. 2011).

  % shock visualization
  To display the shock surface we use the following method. Since we determine
  the shock normal we can assign to each particle with a radio emission above a
  minimum value a small square with the shock normal perpendicular on it. The
  side length of the square is proportional to the SPH smoothing length of the
  particle. The combination of all the small squares represents the shock
  surface, as can be seen in Fig.~\ref{fig:shock-visual}. With a 3D rendering
  the geometry of the shocks can be easily recognized. The simulations allows
  also to study how radio relics evolve. We find that our radio relic is very
  bright right after the first core passage of the cluster merger (left panel)
  and then continuously decreases in brightness (right panel), see
  Fig.~\ref{fig:evolution}.
  \footnote{Animations can be found on \\
           {\tt www.tls-tautenburg.de/research/hoeft/research{\_}struc{\_}shock.html}}

\begin{figure}
\centering
\includegraphics[width=0.7\textwidth]{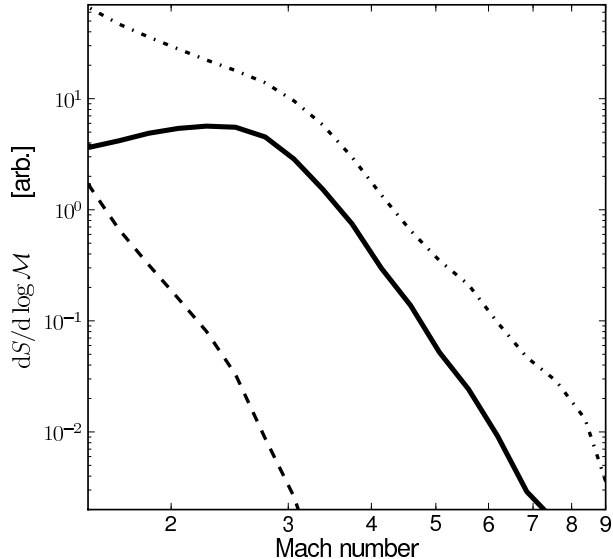}
\caption{
   Distribution of Mach numbers ${\rm d} S / {\rm d} \log {\cal M}$, where $S$
   is the shock surface area. We have selected a section of a spherical shell which
   includes the merger shock front as shown in Fig.~\ref{fig:shock-visual} and
   have determined with our shock finder the Mach number distribution (solid line). 
   From our modeling of X-ray observations we expect a Mach number of 2.5 from the
   temperature jump in the projected emission map. For comparison we show the Mach
   number distribution in the entire cluster up to a radius of 1\:Mpc (dashed line) 
   and up to 3\:Mpc (dash-dotted line).
   }
\label{fig:dSdlogM}
\end{figure}

  % Mach number distribution
  We determine the Mach number distribution ${\rm d} S / {\rm d} \log {\cal M}$
  in the region of the merger shock, see Fig.~\ref{fig:dSdlogM} (solid line).
  The distributions peaks at the Mach number of the shock, 2.5, as obtained from
  analyzing the jump in the X-ray emission. However, the distribution is rather
  broad and becomes almost a power law for ${\cal M} \gtrsim 4$. The Mach number
  varies due to upstream differences in density and temperature. As a
  consequence the radio luminosity varies significantly across the shock surface.

\section{How many relics can be found?}

  %  radio relic probability distribution
  Currently, there are about three dozen galaxy clusters known which show one or
  more radio relic. This allows to start analyzing relics in a statistical way.
  A natural approach would be to determine a `radio luminosity function of shock
  fronts'. More precisely, cosmological simulations should provide a list of
  shocks which then can be populated with relics, similar to populating dark
  matter halos with galaxies. However, as shown above, shock fronts have complex
  geometries and display a broad Mach number distribution. So we cannot simply
  sort shock fronts, e.g. according to their Mach number, and then distribute
  the luminous relics to the shock fronts with the highest Mach numbers. In Nuza
  et al. (2011, in prep.) we developed a different approach: We introduce a {\it
  radio power probability distribution} which give the probability to find a
  relic with a given luminosity for a cluster with given mass and redshift.
  Motivated by simulations we adopt a log-normal distribution, with scaling laws
  depending on cluster mass and redshift. 
  
  % detection probability
  However, several rather bright radio relics have been discovered only
  recently, hence, the flux threshold above which the current list of relics is
  complete would be at least a few hundred mJy, leaving only a few relics in the
  flux-complete list. Instead of using a sharp flux threshold we introduced the
  {\it discovery probability}, which gives a smooth transition between the
  non-discovery and the discovery of relics. The parameters of the discovery
  probability can only be estimated since we do not know how many rather bright
  relics we are not yet detected. In Nuza et al. (2011, in prep.)  we present
  some plausible estimates.

\begin{figure}
\centering
\includegraphics[width=0.7 \textwidth]{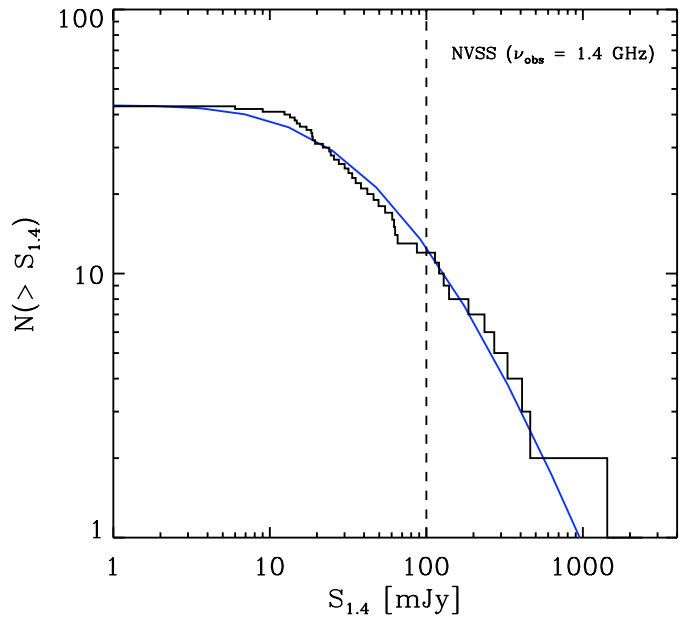}
\caption{
  Number counts $N(>S_{1.4})$, where $S_{1.4}$ is the radio flux of one or more
  relics within one cluster. In Nuza et al. (2011, in prep.) we compile a list
  of all clusters for which one or more relics are known. The stepped curve
  shows the cumulative number counts, while the solid line shows the fit
  obtained from {\it radio power probability distribution}, after normalizing to
  the total number of known relics.
  }
\label{fig:NVSS}
\end{figure}

  % Predictions
  It is impossible to predict the number of observable relics purely based on
  simulations, mainly because we can only speculate about the electron
  acceleration efficiency. We therefore normalize the predicted number counts
  with the number of observed relics, see Fig.~\ref{fig:NVSS}. As a result, we
  can infer expected number counts for future low frequency surveys. In Nuza et
  al. (2011, in prep.) we show that both, the LOFAR-120~MHz-Tier~1 survey and
  the proposed WODAN survey with the upcoming APERTIF sensors at WSRT, have the
  potential to detect of the order of 1000 radio relics. Crucial for actually
  confirming the relics candidates is that the corresponding galaxy clusters
  need to be identified, since most of the low luminosity relics reside in
  clusters with low X-ray brightness. More precisely, we predict that 50~\% of
  the relics with a flux above 1~mJy at 1.4~GHz reside in clusters with an X-ray
  flux below $10^{-12} ~ \rm erg \: s^{-1} \: cm^{-2}$.

  % Comparison with X-ray
  In a recent work van Weeren et al. (2011, submitted) selected 544 clusters
  from the NORAS and the REFLEX cluster samples with an X-ray flux above $3
  \times 10^{-12} ~ \rm erg \: s^{-1} \: cm^{-2}$ which are located outside the
  galactic plane. For 17 of the clusters a radio relic has been reported in the
  literature. Based on this cluster sample van Weeren et al. (2011) find
  tentative evidence that the fraction of clusters which host a relic increases
  with both, cluster X-ray luminosity and redshift. The large relic samples
  expected from the upcoming low frequency surveys will allow tight
  constraints on such behaviour.

\section{Summary}

  % model composed of two parts
  Relics have been found in about three dozen clusters. They are believed to
  originate from merger shocks in clusters. One plausible scenario for relic
  formation is that a small fraction of thermal electrons gets accelerated at
  the shock front to relativistic energies. Bound to the gas flow by small
  gyroradii the relativistic electrons advect downstream and emit synchrotron
  radiation. Hydrodynamical cosmological simulations allow to study merger
  shocks in detail. To study as well the formation of radio relics a sub-grid
  model for both, the electron acceleration and the downstream cooling is
  needed. The relic in CIZA~2242 shows that the intrinsic width is of the order
  of 10 to 50~kpc. Hence, current cosmological simulations are close to
  spatially resolving the downstream cooling, but in our case a sub-grid module
  for it is justified.

  % shock surface is wiggled ..
  We presented a high resolution simulation of a cluster merger. The resulting
  shock front shows a complex surface: There are many `wiggles' on the surface
  which are caused by the inhomogeneity of the ICM. Temperature and density
  variations lead to a not uniform shock speed. The upstream temperature affects
  the Mach number. We determined the Mach number distribution of the shock front
  and find that it peaks at the value expected from the X-ray jump, namely
  2.5. Moreover, we find a rather broad distribution of Mach numbers, up to 4.5.
  Since the actual radio luminosity is believed to vary strongly for Mach
  numbers from 2 to 5 the luminosity is very inhomogeneously distributed across
  the shock surface.

  % predictions
  It would be nice to develop a `radio luminosity function of shock fronts'.
  However, the high resolution merger simulation illustrates that the shock
  fronts can be patchy, with low surface brightness regions between more
  luminous ones. As a consequence, in Nuza et al. (2011, in prep.) we introduce
  the {\it radio power probability distribution}, which relates the
  radio power of relics in a cluster to the cluster mass instead of the shock
  surface. Since the actual electron acceleration efficiency is not known we
  normalize the probability distribution by the number of observed relic. As a
  result we find the LOFAR-120~MHz-Tier 1 survey could detect more than a thousand
  relics.
  \\

  {\sc Acknowledgements.}  
  SEN and MH thank Gustavo Yepes for providing access to the {\it Mare Nostrum
  Universe}. Simulations were
  carried out at the BSC-CNS (Barcelona, Spain) and at the Leibniz-Rechenzentrum
  (Munich, Germany). 
  MH and MB acknowledge support by the research group
  FOR 1254 `Magnetisation of Interstellar and Intergalactic
  Media: The Prospects of Low-Frequency Radio Observations'
  founded by the Deutsche Forschungsgemeinschaft.

\end{document}